# Mitigating the contact resistance limitation of cavitated fine line Ag paste by Laser-Enhanced Contact Optimization

Donald Intal[1], Abasifreke Ebong[1], Vijay Upadhyaya[2], Brian Rounsaville[2], Ajeet Rohatgi[2], Dana Hankey[3], Marshall Tibbetts[3]

[1]University of North Carolina at Charlotte, Charlotte, NC 28223, USA

[2]Georgia Institute of Technology, Atlanta, Georgia 30332, USA

[3]ACI Materials Inc., Goleta, California 93117, USA

*Abstract*—Cavitation-assisted Ag paste is a promising route for fine-line, low-silver metallization in silicon solar cells because it improves paste dispersion, extends shelf life, and reduces Ag consumption, but matching the contact performance of commercial pastes remains a challenge. Here, cavitated paste was evaluated on PERC solar cells at peak firing temperatures of 720, 740, 750, and 762 °C, with and without laser-enhanced contact optimization (LECO). The results show a clear firing window: 720 and 740 °C produced high series resistance and reduced fill factor, 750 °C gave the best pre-LECO performance, and 762 °C showed additional electrical limitations with only limited LECO benefit. LECO selectively recovered the under-activated states, increasing fill factor from 76.8 to 80.2% at 720 °C and from 76.7 to 79.8% at 740 °C. Electroluminescence and conductive AFM further indicated improved current collection and stronger localized conduction after LECO. These results show that cavitated paste performance is governed primarily by a shifted contact-formation window, and that firing optimization combined with LECO provides a practical route to retain the fine-line advantage while improving electrical performance.

*Keywords: cavitated ag paste, silver paste, silicon solar cells, crystalline silicon photovoltaics, fine-line metallization, screen-printed metallization, contact formation, contact resistivity, firing temperature, laser-enhanced contact optimization, electroluminescence, conductive atomic force microscopy, fill factor*

## I. Introduction

Front-side metallization remains one of the most performance-critical process modules in crystalline silicon solar cells because it simultaneously determines optical shading, lateral current transport, and metal-silicon contact formation [1]. Narrow screen-printed Ag fingers are desirable because they reduce shading and metal-induced recombination, but this advantage is only realized when low line resistance and low contact resistivity are maintained [1], [2]. In practice, metallization is therefore not simply a printing problem. It is a coupled materials and interface problem in which paste composition, microstructure, and thermal processing together determine whether a fine printed grid can also deliver high fill factor [1].

For conventional fire-through Ag contacts, low-ohmic contact formation develops during rapid firing through a sequence of frit softening, dielectric etching, Ag dissolution and transport, and interfacial precipitation [2]. In the widely cited mechanism reported by Fields et al., the molten frit wets the Ag-SiNx interface above about 500 °C, PbO-containing frit components etch and penetrate the SiNx layer between about 500 and 650 °C, and above about 650 °C Ag dissolves into the molten glass, diffuses toward the Si surface, and subsequently precipitates as metallic Ag and distributed nanocrystals during cooling [2]. The resulting contact is therefore governed by the density and continuity of local conductive pathways across a thin glass-containing interfacial region rather than by bulk silver conductivity alone [3], [4], [5].

Because these reactions are tightly coupled to the thermal budget, screen-printed Ag contacts exhibit a pronounced firing window [6], [7]. Under-firing generally leaves contact activation incomplete, whereas excessive firing can alter the interfacial glass layer, intensify local damage to passivation, or shift the balance between contact formation and recombination loss [6], [7]. Earlier microstructural studies have shown a clear correlation between firing-dependent interfacial morphology and specific contact resistance, reinforcing that optimum firing is paste-specific and cannot be assumed to transfer from one metallization system to another [6], [7]. This sensitivity is especially relevant when the paste microstructure itself is deliberately modified [8].

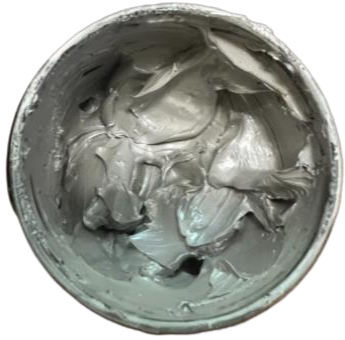

*Figure 1: Photograph of the cavitated Ag paste showing retained visual uniformity without macroscopic phase separation after long-term storage from 2021 to 2026, consistent with improved shelf stability of the cavitation-assisted formulation [9], [10].*

Our previous work introduced cavitation technology as a route to modify Ag paste microstructure to nanostructures for finer-line metallization [9], [10]. That study showed that cavitation improved paste dispersion, enabled longer shelve life without constantly rotating, and aided finer gridlines, thereby supporting reduced silver usage and competitive device efficiency [9], [10]. As shown in Figure 1, the cavitated paste prepared in 2021 remained visually uniform without observable phase separation in 2026, providing direct macroscopic evidence of the extended shelf stability associated with the cavitation-assisted formulation [9], [10]. At the same time, the cavitated paste exhibited somewhat higher contact resistance and lower fill factor relative to the conventional reference [10]. The central unresolved issue from that work was therefore not whether cavitation can improve printability, but whether the resulting paste can also form sufficiently active electrical contacts at the Ag-Si interface [10]. The conceptual basis of this follow-up study is summarized in Figure 2, which contrasts the fine-line metallization advantage of the cavitated paste with the unresolved contact-limited loss identified in the previous work.

This issue leads to the next scientific question: does cavitation shift the effective contact formation window? That possibility is physically plausible. Cavitation changes particle size distribution and dispersion state, and those changes are expected to affect sintering kinetics, pore evolution, and frit-mediated transport during rapid firing [8], [10]. If the balance between densification of the printed finger and activation of the buried contact is altered, a firing condition that is suitable for a conventional paste may leave the cavitated paste in a partially activated state, with too few or too discontinuous local current pathways. In that case, the fill factor penalty would not represent an intrinsic limitation of cavitation itself, but rather a process-window mismatch that should be recoverable through contact-focused process optimization. As illustrated schematically in Figure 2, the central question is whether firing optimization and LECO can recover the incompletely activated contact network while preserving the fine-line advantage of the cavitated paste.

Laser-enhanced contact optimization, or LECO, provides a particularly useful framework for testing that hypothesis. LECO has been reported as a post-firing treatment in which a localized laser scan is applied under reverse bias, originally to recover underfired contacts and reduce metal-semiconductor contact resistivity without requiring a globally higher firing temperature [11], [12]. Fellmeth et al. showed that LECO can substantially reduce contact resistivity in underfired cells and relax the dependence of fill factor on peak firing temperature, while Großer et al. provided microstructural evidence that LECO generates numerous microscopic contacts at the buried Ag-Si interface and is associated with local Ag-Si interdiffusion [11], [13]. For the present study, LECO is therefore not only a recovery step but also a mechanistic probe. If the cavitated paste is indeed contact limited because of incomplete activation during firing,

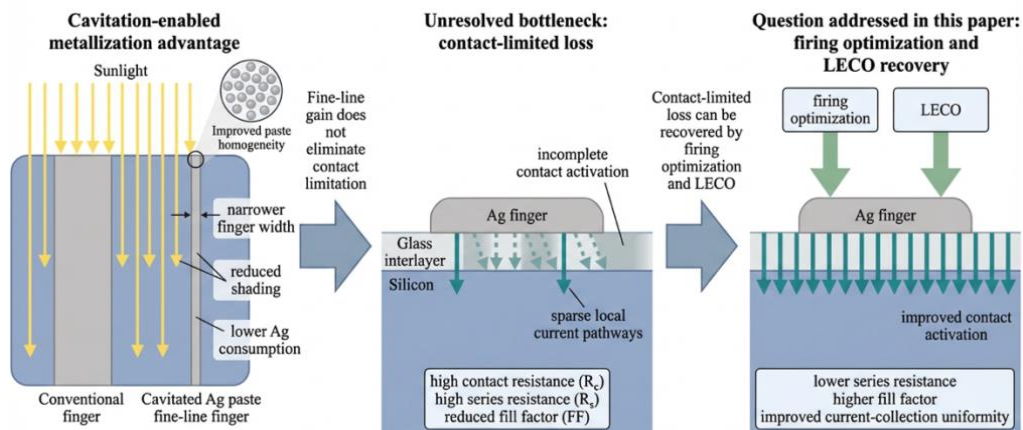

*Figure 2: Conceptual schematics of the follow-up study. Cavitation enables fine-line Ag metallization with reduced shading and lower Ag consumption, but incomplete contact activation leaves sparse current pathways and causes contact-limited loss. This work examines whether firing optimization and LECO can recover the contact network and restore fill factor.*

LECO should preferentially improve those marginal states.

Demonstrating that mechanism requires diagnostics across multiple length scales. Device-scale current-voltage measurements identify whether fill factor and efficiency are limited by resistive loss, but they do not reveal where the limitation originates. Electroluminescence is valuable because it can spatially resolve local series-resistance-related nonuniformity, and EL-based methods have long been used to visualize resistive contrast in silicon solar cells [14]. At the local scale, microscopic electrical studies have shown that silver crystals and discrete contact sites at the interface play a major role in current transport, while nanoscale and cross-sectional resistance imaging has linked Ag-grid-related EL features to local loss of electrical contact [3], [5], [15]. Against this background, conductive AFM on representative gridline regions is a rational addition to the present study because it can probe local current heterogeneity and test whether optimized firing or LECO increases the density or continuity of active conduction pathways, even if it does not directly image the entire buried interface [15].

Accordingly, the objective of this work is to determine whether cavitation-enabled Ag paste exhibits a shifted firing-dependent contact formation window and whether LECO can recover the associated contact-limited performance loss. We combine current-voltage analysis across a firing-temperature matrix with pre- and post-LECO comparison, electroluminescence imaging, and local current mapping by conductive AFM to construct a process-structure-property-performance
relationship for cavitated Ag metallization. We hypothesize that non-optimized firing leaves the cavitated paste with a spatially incomplete network of electrically active contact pathways, and that LECO partially restores fill factor by increasing the number or continuity of those pathways while preserving the fine-line advantage of the cavitated paste.

## II. Experimental

### A. Ag paste preparation and solar cell fabrication

The cavitated Ag paste used in this study was prepared following the cavitation-assisted route reported in our previous work [10]. In brief, the Ag powder and organic vehicle were processed under cavitation conditions designed to improve particle dispersion and paste homogeneity. The resulting paste was used as the front-side metallization material for crystalline silicon solar cells. A conventional reference paste was also included where indicated for comparison.

Solar cells were fabricated on industrial G1 PERC wafers with a size of 158.75 mm x 158.75 mm and a base resistivity of 1±0.05 Ω·cm. The front-side paste was deposited by screen printing using a 40 μm opening screen under identical printing conditions for all groups. The printed wafers were dried at 200 °C for 1 minute prior to fast firing in an infrared belt furnace. Unless otherwise stated, all processing steps other than peak firing temperature were kept constant to isolate the effect of firing on contact formation and device performance.

### B. Firing temperature matrix

To evaluate the firing dependence of the cavitated paste, solar cells were processed over a defined peak firing temperature window of 720-760 °C with a step size of 10-20 °C. This range was selected to cover underfired, near optimum, and over-fired conditions. For each firing condition, 10 cells were fabricated and measured. The belt speed, zone configuration, and drying conditions were kept constant throughout the study, while only the peak set temperature was varied.

This experimental design was used to determine whether the cavitated paste exhibits a shifted or narrowed contact formation window relative to the baseline condition used in the previous study.

### C. LECO treatment

After initial electrical characterization, selected cells were subjected to laser-enhanced contact optimization (LECO by Cell Engineering). LECO was performed under 18% power and a reverse-bias condition of 15V, following the process window available for the cell architecture used in this work. The same cells were measured before and after LECO to directly evaluate the treatment-induced change in device performance and minimize sample-to-sample variation.

LECO was applied to cells fired at different peak temperatures so that its effect could be compared across under-activated, near optimum, and over-fired contact states.

### D. IV Characterization

Current-voltage measurements were carried out under standard test conditions using a FCT-650 (Sinton Instruments) at AM1.5G illumination, 1000 W m$^{-2}$, and 25 °C. The open-circuit voltage, short-circuit current, fill factor, and conversion efficiency were extracted for each cell. Series resistance was obtained from the IV analysis using the instrument software or by the standard extraction procedure implemented in the measurement system.

For the LECO study, each cell was measured immediately before and after treatment under identical conditions. The change in fill factor, efficiency, and series resistance was used as the primary indicator of firing-dependent and post-treatment-dependent contact recovery.

## E. Electroluminescence Imaging

Electroluminescence images were acquired using an FCT-650 built in system at an injection condition of 1V and an exposure time of 0.2 s. All EL measurements were performed under identical imaging conditions to allow direct comparison between samples. For the LECO study, paired EL images were recorded from the same cell before and after treatment.

The EL analysis was used to assess the spatial uniformity of current collection and to identify resistive nonuniformity associated with firing condition and post-metallization recovery. Representative images from low, intermediate, and high firing temperatures were selected for comparison. Where applicable, line profiles or normalized intensity distributions were extracted from the EL images to support the qualitative image interpretation.

## F. Conductive Atomic Force Microscopy

Local current transport in representative metallized regions was investigated by conductive atomic force microscopy using an MFP-3D (Asylum Research) equipped with a conductive soft tapping mode, Pt-coated (30 nm platinum) AFM probe (HQ:NSC18/Pt). Measurements were carried out in contact mode under an applied tip bias of 10V with a scan size of 5μm and a scan rate of 1 Hz.

Conductive AFM was performed on selected samples representing for sample before LECO, the same condition after LECO. Current maps were recorded together with topography images to compare local conduction behavior across conditions. The conductive AFM analysis was used to evaluate local current heterogeneity within the printed gridline region and to determine whether firing optimization or LECO increased the density or continuity of electrically active pathways.

# III. Result & Discussion

## A. Firing-window dependence of the electrical performance of cavitated Ag paste cells

The electrical performance of the cavitated Ag paste is strongly dependent on peak firing temperature, indicating that the main limitation is not an intrinsic inability to form a functional contact, but a shifted and relatively narrow contact-formation window. Table 1 shows that the low-temperature side of the window, represented by 720 and 740 °C, is dominated by a pronounced fill-factor penalty together with high series resistance. In contrast, the intermediate condition at 750 °C gives the best overall balance of device parameters, while the highest temperature, 762 °C, no longer improves performance and instead begins to show signs of additional electrical loss. This progression indicates that the dominant problem at low firing temperature is incomplete contact activation.

That interpretation is supported by the relative stability of $V_{OC}$ and $J_{SC}$ across the firing range. $J_{SC}$ changes only slightly, from 41.0 to 41.3 mA cm$^2$, and $V_{OC}$ remains within 670 to 677 mV. By contrast, FF and $R_S$ change much more strongly. Before LECO, FF increases from 76.7 to 80.1% as the firing temperature increases from 740 to 750 °C, while $R_S$ falls from about 1.25 to 0.63 Ω cm$^2$. The low-temperature loss is therefore not consistent with an optics-driven or current-generation-limited problem. Rather, it reflects inefficient current extraction through the front contact.

The 750 °C condition defines the practical optimum within the present firing matrix. At this temperature, the nonLECO cell reaches the highest nonLECO FF and efficiency together with the lowest $R_S$, indicating that contact activation is closest to completion. Raising the peak firing temperature further to 762 °C does not provide additional benefit. Instead, FF and efficiency decline, $V_{OC}$ drops, $R_{SH}$ decreases sharply, and $J_{02}$ increases. This high-temperature behavior is important because it shows that the cavitated paste does not simply require more aggressive firing. Rather, it has a limited operating window in which insufficient firing leaves the contact under-activated, whereas excessive firing begins to introduce junction-related or leakage-related penalties.

*Table 1: IV parameters of cavitated Ag paste solar cells fired at different temperatures before and after LECO.*

| Firing (°C) | Sample ID | VOC (mV) | JSC (mA/cm2) | FF (%) | Eff. (%) | RS (Ω-cm2) | RSH (Ω-cm2) | pFF (%) | n (1 sun) | n (0.1 sun) | J02 (nA/cm2) |
|---|---|---|---|---|---|---|---|---|---|---|---|
| 720 | nonLECO | 675 | 41.2 | 76.8 | 21.4 | 1.254 | 82179 | 83.8 | 0.948 | 1.078 | 2.9 |
| 720 | LECO | 676 | 41.2 | 80.2 | 22.3 | 0.618 | 76945 | 83.7 | 0.976 | 1.088 | 2.9 |
| 740 | nonLECO | 677 | 41.3 | 76.7 | 21.4 | 1.245 | 100153 | 83.9 | 0.953 | 1.079 | 2.9 |
| 740 | LECO | 678 | 41.2 | 79.8 | 22.3 | 0.688 | 95573 | 83.7 | 0.977 | 1.089 | 3.5 |
| 750 | nonLECO | 674 | 41.1 | 80.1 | 22.2 | 0.630 | 70268 | 83.8 | 0.970 | 1.077 | 3.2 |
| 750 | LECO | 674 | 41.1 | 80.8 | 22.4 | 0.499 | 66517 | 83.7 | 0.976 | 1.080 | 3.3 |
| 762 | nonLECO | 670 | 41.0 | 79.2 | 21.8 | 0.770 | 27939 | 83.6 | 0.974 | 1.093 | 3.9 |
| 762 | LECO | 671 | 41.0 | 79.5 | 21.9 | 0.705 | 20598 | 83.5 | 0.975 | 1.099 | 4.0 |

LECO modifies this firing response in a selective manner. The strongest gains occur exactly where the preLECO cells are most resistive. At 720 °C, FF increases from 76.8 to 80.2% and $R_S$ decreases from 1.254 to 0.618 $\Omega$ cm$^2$. At 740 °C, FF increases from 76.7 to 79.8% and $R_S$ decreases from 1.245 to 0.688 $\Omega$ cm$^2$. By contrast, the gains at 750 and 762 °C are much smaller. This selective response is one of the central findings of the study. It shows that LECO is not acting as a uniform performance booster but is preferentially recovering those firing states in which the contact remains only partially activated after firing.

The same conclusion is evident in the efficiency trend. Before LECO, efficiency rises from 21.4% at 720 and 740 °C to 22.2% at 750 °C, then falls again to 21.8% at 762 °C. After LECO, the 720 and 740 °C cells recover to 22.3%, approaching the 22.4% obtained at 750 °C. LECO therefore compresses the low-temperature side of the firing window by rescuing the under-activated states. In practical terms, the cavitated paste is not uniformly contact-limited across all firing temperatures. Instead, the contact limitation is strongest at the low-temperature side of the process window and becomes highly recoverable after LECO.

## B. Selective fill-factor recovery by LECO through resistive-loss reduction

Once the firing window is established, the next question is whether LECO mainly reduces transport loss or significantly changes intrinsic diode quality. The data indicate that the recovery is dominated by resistive-loss reduction. The clearest evidence is the behavior of pFF. Across all firing temperatures and treatment states, pFF remains nearly constant, varying only from 83.5 to 83.9%. By contrast, the measured FF changes strongly with both firing and LECO. This combination of nearly unchanged pFF and strongly recovered FF is the expected electrical signature of a transport-dominated improvement.

This interpretation becomes even clearer when the pFF minus FF gap is considered. At 720 and 740 °C, the gap is large before LECO and narrows strongly after LECO treatment. At 750 °C, where the contact is already close to optimum, the gap is much smaller to begin with and only improves modestly after LECO. At 762 °C, the change is limited. This behavior tracks the $R_S$ trend closely. The same cells that show the largest LECO-induced reduction in $R_S$ also show the largest narrowing of the pFF minus FF gap. The fill-factor gain is therefore best interpreted as recovery of current transport rather than improvement of the underlying diode limit.

The stability of $V_{OC}$ and $J_{SC}$ supports the same conclusion. LECO changes $V_{OC}$ by only about 0 to 1 mV and leaves $J_{SC}$ essentially unchanged. These shifts are too small to explain the several-percentage-point increase in FF at low firing temperature. As a result, the efficiency gain follows FF much more closely than either $V_{OC}$ or $J_{SC}$. In physical terms, LECO is not creating a new current-generation regime or a major increase in junction voltage. It is allowing the existing current to be extracted more efficiently through the front-contact system.

The diode-related fitting parameters reinforce this interpretation, although they should be handled cautiously. The extracted n values at 1 sun and 0.1 sun vary only modestly with firing and LECO when compared with the large changes in FF and $R_S$. Likewise, $J_{02}$ changes only slightly over most of the matrix, although it rises somewhat at the high-temperature side. Because the absolute meaning of these fitted parameters depends on the extraction routine and on how series and shunt contributions are incorporated into the fit, they are most useful here as relative indicators rather than as independent mechanistic proof. Their limited variation nevertheless remains important. It shows that the dominant LECO effect is not a major change in the diode itself, but a change in how effectively the front metallization is electrically coupled to the junction.

A further point helps explain the magnitude of the electrical gain. The measured $R_S$ is not purely a contact term. It includes contributions from the front finger, the emitter, the contact interface, the bulk, and the rear side. When LECO improves effective contact, current transfer becomes less crowded near the finger perimeter and more uniformly distributed across the contacted region. As a result, the apparent reduction in $R_S$ can include both interfacial and emitter-coupled contributions. This is important in the present data because it explains why the FF recovery can be large even if the extracted average contact resistivity does not change dramatically.

The 762 °C condition again acts as a useful boundary case. LECO still lowers $R_S$ slightly, but $R_{SH}$ decreases further, $J_{02}$ remains elevated, and efficiency improves only marginally. This indicates that once the firing condition moves beyond the practical optimum, further performance limitation is no longer governed mainly by incomplete contact activation. The modest LECO benefit at 762 °C therefore strengthens the interpretation that the large low-temperature recovery is specifically transport and contact related.

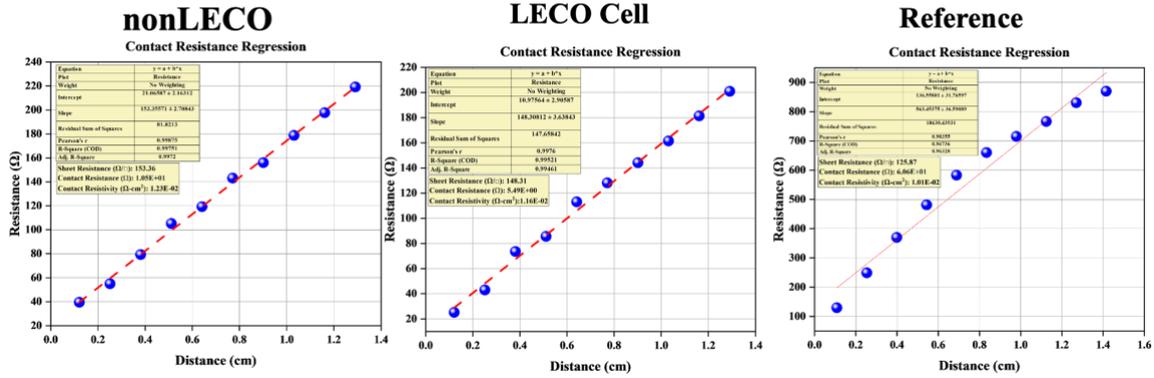

*Figure 3: Contact-resistance regressions of representative cavitated Ag paste contacts before and after LECO.*

## C. Evolution of contact resistance relative to industrial reference paste

The contact-resistance regressions in Figure 3 provide a more direct, although still averaged, view of the contact. The non-LECO cavitated cell gives an extracted sheet resistance of about 153.36 Ω/sq, a contact resistance of about $1.05 \times 10^1$ Ω, and a contact resistivity of about $1.23 \times 10^{-2}$ Ω-cm². After LECO, the extracted sheet resistance decreases slightly to about 148.31 Ω/sq and the extracted contact resistance decreases more strongly to about $5.49 \times 10^0$ Ω, while the extracted contact resistivity decreases only modestly to about $1.16 \times 10^{-2}$ Ω-cm². By comparison, the industrial reference paste gives a somewhat lower extracted contact resistivity of about $1.01 \times 10^{-2}$ Ω-cm². Thus, all three samples remain within the same $10^{-2}$ Ω-cm² regime, and LECO shifts the cavitated paste toward the industrial reference without fully converging to it.

This modest change in averaged extracted contact resistivity contrasts with the much larger recovery observed in Table 1, particularly the strong reduction in series resistance and the more than 3% absolute gain in fill factor at low firing temperature. However, this apparent mismatch is not necessarily contradictory. In practical TLM analysis of finished solar-cell strips, the extracted contact resistivity is an averaged effective quantity and may not fully capture the small fraction of local bottlenecks that dominate cell-level current transport. Within that framework, the present results suggest that LECO improves the practical electrical quality of the contact more strongly than is reflected by the small change in averaged extracted ρc. This interpretation is supported by the much larger decrease in the extracted contact-resistance term, the strong reduction in cell-level series resistance, and the substantial recovery in fill factor. A physically consistent explanation is that LECO acts primarily on the electrically limiting fraction of the contact network. If the most resistive local pathways are activated first, device-level current transport can improve substantially even while the strip-averaged ρc remains within the same decade.

## D. Electroluminescence evidence of improved current-collection uniformity

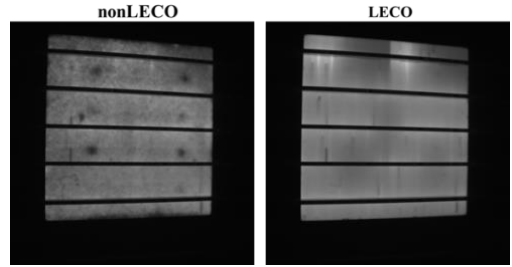

*Figure 4: Electroluminescence images of representative cavitated Ag paste solar cells before and after LECO.*

The EL images in Figure 4 provide the spatial counterpart to the electrical trends in Table 1. The nonLECO cell shows a darker and more mottled emission pattern, with obvious local dark features and broader nonuniformity across the active area. After LECO, the cell becomes visibly brighter and much more homogeneous. The important point is not simply that the treated image looks better, but that the spatial change is fully aligned with the electrical recovery already established in the IV data.

A more uniform EL response implies that the local junction voltage under forward bias is distributed more evenly across the cell. In the present case, that observation is consistent with a reduction in lateral transport loss and improved coupling between the front metallization and the emitter. The EL result therefore adds mesoscale evidence to the device-level conclusion that LECO does not merely improve one isolated location. Rather, it improves current collection across a broad fraction of the active area.

This spatial perspective is especially valuable because it supports the interpretation of LECO as recovery of under-activated contacts rather than a purely numerical change in extracted fit parameters.

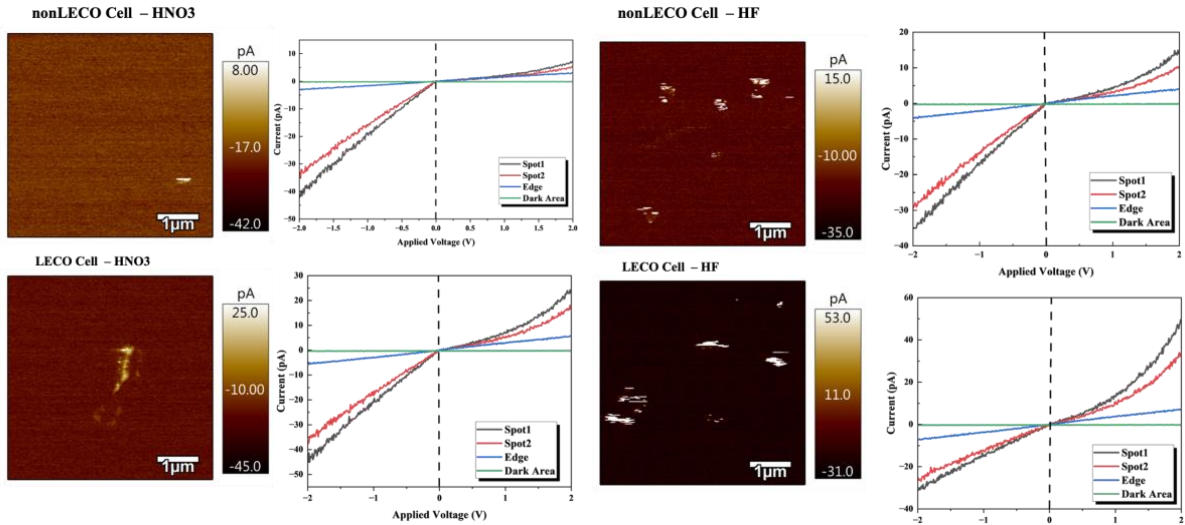

*Figure 5: Conductive AFM current-voltage characteristics of edge, dark area, and localized conductive spots on non-LECO and LECO surfaces after HNO3 and HF treatment.*

The low-temperature cells show the largest recovery in FF and $R_S$, and the EL images show that such recovery is accompanied by a more uniform emission pattern. In that sense, Figure 4 is not a separate visual add-on to the electrical data. It is part of the same physical picture, showing that the transport bottleneck identified in Table 1 has a real spatial signature at the cell scale.

## E. Conductive AFM: localized pathway activation

The conductive AFM measurements in Figure 5 provide local-scale evidence that complements the cell-level electrical and electroluminescence results. After sequential HNO3 and HF treatments, clear spatial contrast is observed among the analyzed regions, indicating that the revealed surface is electronically heterogeneous rather than uniformly conductive. In all four datasets, the edge region shows the most nearly linear and symmetric current-voltage response, which is consistent with ohmic-like transport under the applied bias range. By contrast, the dark region remains close to the current floor in all conditions, indicating that it is largely insulating or electronically inactive. The main electrical activity is therefore concentrated at discrete conductive spots rather than distributed uniformly across the revealed surface.

This interpretation is supported by the behavior of Spot1 and Spot2 in Figure 5. In both the non-LECO and LECO samples, these spots show distinctly nonlinear and asymmetric current-voltage curves, indicating barrier-limited transport rather than simple resistive conduction. The spots therefore do not behave like ordinary metallic pathways. Instead, they represent localized conduction sites whose electrical response is still influenced by an interfacial barrier at the tip-sample junction or in the near-surface region exposed by chemical revealing.

For the non-LECO surface after HNO3 etching, Spot1 and Spot2 show strong rectification, with much larger current under negative bias than under positive bias. This behavior indicates that charge injections remain strongly polarity dependent and that the conductive sites are still in a partially blocked state. After the additional HF treatment, both spots become more conductive, especially at positive bias, and the rectification becomes weaker. The most direct interpretation is that HF removes or thins a near-surface barrier that remains after $HNO_3$ etching, making the conductive sites more electronically accessible even though their response is still not fully ohmic.

The LECO surface follows the same overall trend, but with a much stronger response. After $HNO_3$ etching, the spot currents are already higher than in the non-LECO case, indicating that the LECO-treated material contains more electronically active conductive sites. After the additional HF treatment, the increase in spot current becomes even more pronounced, and the polarity dependence changes substantially. In the LECO plus HF condition, the positive-bias current becomes comparable to or greater than the negative-bias current, showing that the original barrier asymmetry has been strongly modified. This behavior is consistent with HF removing an oxide-like or chemically altered surface layer that previously controlled the injection process and exposing a more active underlying conduction pathway.

Taken together, the c-AFM data show that $HNO_3$ alone leaves the conductive spots in a partially barrier-limited state, whereas HF makes those spots more electronically accessible by reducing interfacial resistance or barrier thickness. The fact that this change is much larger for LECO than the non-LECO is particularly important. It implies that LECO does not simply increase average conductivity uniformly. Instead, it produces a surface or subsurface structure that can support much stronger localized conduction once the blocking layer is removed. The persistent ohmic-like response at the edge and the weak activity in the dark region further show that different regions are governed by different transport mechanisms.

## IV. Conclusion

This study shows that the electrical penalty of the cavitated Ag paste arises mainly from a shifted and relatively narrow contact-formation window rather than from an intrinsic loss of metallization capability. Low firing temperatures were limited by high series resistance and reduced fill factor, while 750 °C provided the best overall device performance. LECO selectively recovered the under-activated states, producing the largest improvements at 720 and 740 °C and only limited benefit near or beyond the optimum firing condition. The combined IV, contact-resistance, electroluminescence, and conductive AFM results indicate that this recovery is transport-related and is associated with strengthening the electrically active fraction of a heterogeneous contact network. These findings show that the fine-line advantage of the cavitated paste can be retained when contact activation is properly controlled through firing optimization and post-metallization treatment.